\documentclass[twocolumn,aps,prl,amsmath,showpacs]{revtex4-1}
\usepackage{graphicx}
\usepackage{dcolumn}%
\usepackage{bm}%
\usepackage{amssymb}
\usepackage{amsmath}
\usepackage{color}

\def\be{\begin{equation}}
\def\ee{\end{equation}}

\begin{document}

\title{Shear-driven solidification of dilute colloidal suspensions}
\author{Alessio Zaccone $^{1}$, Daniele Gentili $^{1}$, Hua Wu $^{1}$, Massimo
Morbidelli$^{1}$, and Emanuela Del Gado$^{2}$}
\affiliation{${}^1$Chemistry and
Applied Biosciences, ETH Z\"urich, CH-8093 Z\"urich, Switzerland}
\affiliation{${}^2$Microstructure and Rheology, Institute for Building Materials, ETH Z\"urich, CH-8093 Z\"urich,
Switzerland}
\date{\today}
\begin{abstract}
We show that the shear-induced solidification of dilute charge-stabilized (DLVO) colloids is due to the interplay between the shear-induced formation and breakage of large non-Brownian clusters. While
their size is limited by breakage, their
number density increases with the shearing-time.
Upon flow cessation, the dense packing of clusters interconnects
into a rigid state by means of {\it grainy} bonds, each involving a
large number of primary colloidal bonds. The emerging picture of shear-driven
solidification in dilute colloidal suspensions combines 
the gelation of Brownian systems with the jamming of athermal systems.
\end{abstract}

\maketitle

Shear-driven solidification of diluted colloidal suspensions has
dramatic impact on their applications, from
industrial polymer production to natural
microfluidic devices 
\cite{russel,bausch},  and is a prototype of non-equilibrium transitions.
If interparticle interactions are purely attractive,
the applied shear stress may break down 
aggregates and fluidize the material
\cite{vermant,liu}. However, the
colloidal particles are often stabilized by electrostatics
\cite{russel}, with no tendency to aggregate at rest, and high shear rates may ultimately promote
aggregation in competition with the electrostatics.
The interplay between these two tendencies may lead to persistent structures \cite{wu} and
this
shear-induced aggregation might be as dramatic as a
complete solidification of even diluted suspensions,
hence seriously affecting the material and rheological properties. Although this is a widely reported
phenomenon in both artificial and living systems\cite{bausch,wu,vasudevan_wunderlich}, there is little understanding of the solidification mechanism. This is due to the difficulty to monitor the system with real-space optics
or scattering techniques at high shear rates\cite{wu,vasudevan_wunderlich}.
To overcome these obstacles,
we have designed an experimental protocol
exploiting
charge-stabilized colloidal particles
which
interact via a typical DLVO potential \cite{russel} where an
energy barrier coexists with a deep, shorter-ranged
attractive well. In our setup, shear-induced aggregation can
be monitored in a fully controlled way, allowing for the
first time to
rationalize the effect of the shear stress from an initially
dilute suspension to the final solid. By combining
light scattering, rheology, and microscopy data we formulate a
model for the solidification mechanism.
The effective packing fraction of the aggregates formed under shear
increases with time. Upon
flow cessation their dense packing is progressively frozen into a
rigid structure by the formation of {\it grainy} (i.e., multiple) colloidal bonds which are responsible for
the fairly high shear moduli observed.

{\it Experiments.} The system consists of colloidal particles
at a fixed colloid volume fraction of $\phi\simeq 0.21$. For the effective DLVO interaction
\cite{russel}, the attractive well depth is $\simeq 40 k_{B}T$ and the
repulsive barrier $\simeq 60 k_{B}T$. The colloidal particles are
surfactant-free polystyrene-acrylate latex spheres charge-stabilized
by the charged groups of the initiator \cite{zacJPCB}. The nearly
monodisperse particles have mean radius $a \simeq 60 nm$ as determined
from both dynamic and static light scattering (using a
BI-200 SM goniometer system, Brookhaven Instruments, NY). $17 mM$ of
NaCl were added to weakly screen the electrostatic interaction
barrier, characterized by Zeta-potential measurements.
A strain-controlled
ARES rheometer (Advanced Rheometric Expansion System, TA
Instruments, Germany) with Couette geometry has been employed
to induce the shear flow under shear-rate control
and to measure the rheological
properties. We have initially sheared  the
system for a varying time $\tau_{1}$  at a fixed shear-rate ($1700 s^{-1}$).
For each $\tau_{1}$, upon flow cessation,
we sampled the shear cell and
analyzed the system by laser light scattering (LLS) using a
small-angle light scattering Mastersizer 2000 instrument (Malvern,
U.K). The LLS analysis can be done off-line since the aggregation
does not evolve on the time scale of the analysis in the absence of
shear. The samples were diluted at $\dot{\gamma}=0$ to such an
extent that multiple light scattering does not affect the
measurement.
After each $\tau_{1}$ we have also performed rheological shear-sweep and frequency-sweep tests.
Since the height of the energy barrier for bond-breaking is
$ \sim 100
k_{B}T$, the clusters are mechanically stable even under the
fairly high shear rate of the Couette cell,
hence it is clear that
they cannot de-aggregate during the off-line analysis. From the scattered
intensity $\emph{I}(\emph{q})$
we obtain the average
structure factor of the aggregates
$\langle\emph{S}(\emph{q})\rangle$ \cite{dimon} (not to be confused
with the %
one of the whole suspension) present in the
system at the time of sampling.

{\it Results and discussion.} The starting point of our analysis is
the time evolution of the systems sheared at a constant shear rate $\dot{\gamma}= 1700s^{-1}$.
The viscosity shows a sharp upturn in time,
after an induction delay due to the activation barrier in the
microscopic aggregation kinetics between two colloidal particles \cite{alessio1}.
\begin{figure}
\includegraphics[width=1.0\linewidth]{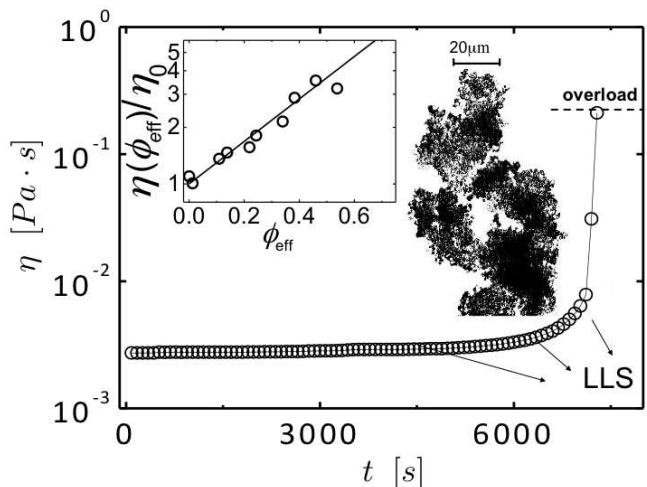}
\vspace{-0.8cm}
\caption{Main frame: Time evolution of the viscosity $\eta$ at $\dot{\gamma}= 1700s^{-1}$ an $\phi\simeq 0.21$.
Insets: left - $\eta/\eta_{0}$ as a function of $\phi_{eff}$
from experiments (circles) and the extended
Einstein formula (line); right -
Optical micrograph upon flow cessation.
} \label{fig1}
\end{figure}
The barrier for shear-induced aggregation, in fact, decreases exponentially upon
increasing the shear rate
and vanishes upon reaching a certain cluster size above which the aggregation kinetics is much
faster and goes with the cube of the cluster size \cite{alessio1}.
Hence a kind of self-accelerated kinetics, leading
to larger and larger aggregates, is expected once clusters of a
critical size are formed. On the other hand, the action of shear is
also well known to cause fragmentation, which should ultimately lead
to a maximal cluster size, dictated by the mechanical balance
between the stress imposed by the flow on the cluster and its
mechanical response. Nevertheless we observe (Fig.\ref{fig1}, main frame) that
the viscosity continues to increase until the instrument stress overload
threshold is reached.
We have stopped the shearing  
after
different time duration $\tau_{1}$ ({\it pre-shearing} times).
For each $\tau_{1}$ we have used optical microscopy and LLS to investigate the
structure of the resulting suspension. An optical micrograph upon flow cessation at the largest $\tau_{1}$ is shown in
the inset (right) of Fig.\ref{fig1}.
From the small angle light scattering analysis \cite{suppl}, it is evident that already after a short
time the size distribution of aggregates is strongly bimodal (i.e.,
primary particles coexist with large aggregates). The Guinier plot gives a
typical aggregate radius, resulting from the competition between
aggregation and breakage, $R_{g}\simeq35 \pm 3\mu m$ which
remains constant with time \cite{suppl}, indicating
that the aggregation under shear rapidly leads to an optimal cluster
size.
\begin{figure}
\includegraphics[width=1.0\linewidth]{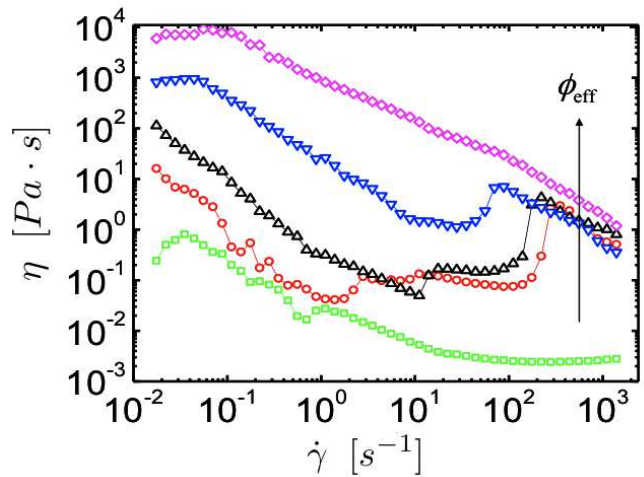}
\vspace{-0.8cm}
\caption{(color online) Shear-sweep curves obtained after pre-shearing at
$\dot{\gamma}=1700s^{-1}$ for
$\tau_{1}=4054 s, 7420 s, 7500 s, 7520 s$ and $7555 s$ (from bottom to top).
} \label{fig2}
\end{figure}
From the power-law regime of the scattering curves we extract
the fractal dimension $d_{f}$ of the large clusters: as a
function of $\tau_{1}$ \cite{suppl}
$d_{f}$ also rapidly reaches a plateau at $\simeq 2.7 \pm 0.1$, a
fairly high value which is typical of shear-driven
aggregation \cite{warren} where breakage events and
restructuring induced by flow stresses combine 
\cite{becker}. We have 
measured the fraction $\chi(t)$ of
primary colloidal particles converted to clusters and found that it
steeply increases 
\cite{suppl}.
The emerging picture is that under shear the system is constantly
generating new clusters with approximately the same size $R_{g}$
and fractal dimesion $d_{f}$. The fast
increase in the number density of clusters which are fractal and
porous is associated with a rapid increase of their effective
packing fraction: this could be the reason for the sharp non-linear increase
of the viscosity in Fig.\ref{fig1}.
To test this hypothesis, we have estimated the clusters effective packing
fraction $\phi_{eff}$ through the relation $\phi_{eff} (t) = \chi(t)
\phi k^{-1} (R_{g}/a)^{3-d_{f}}$, where $k$ is a geometric prefactor
close to one and $\phi$ is the initial volume fraction of primary
particles \cite{zacthesis}. In the inset (left) of Fig.\ref{fig1}
we plot the viscosity $\eta (t)$ (circles), normalized by
$\eta_{0}$, as a function of $\phi_{eff}(t)$. We have also
calculated the high shear viscosity of an equivalent suspension of
hard spheres of the same linear size at the same volume fraction
$\phi_{eff}(t)$, using the Einstein formula properly extended to
high $\phi_{eff}(t)$ \cite{larson} which gives $\eta
\simeq \eta_{0}\exp{(5\phi_{eff}(t)/2)}$ (the full line in the same
inset). The agreement indicates that the increase of the
viscosity under shear in Fig.\ref{fig1} can be ascribed indeed to the
increasing packing fraction of the
clusters, which hydrodynamically behave as hard spheres
due to the fairly high $d_{f}$ \cite{degennes}):
the initial dilute colloidal suspension has
changed, under shear, into a suspension of non-Brownian
aggregates whose packing fraction increases with the shearing time.

After each of the pre-shearing times, we
also perform a shear-sweep experiment where $\dot{\gamma}$ is varied but
kept below $\dot{\gamma}=1700s^{-1}$. This guarantees that the
aggregates formed during the pre-shearing are mechanically stable
and do not break up during the shear-sweep \cite{alessio2,moussa}.
The data are plotted in Fig.\ref{fig2}. The curves correspond to
$\tau_{1}= 4054 s, 7420s, 7500s, 7520$ and $7520 s$ from bottom to
top, and to $\phi_{eff}(\tau_{1})$ increasing accordingly.
\begin{figure}
\includegraphics[width=1.0\linewidth]{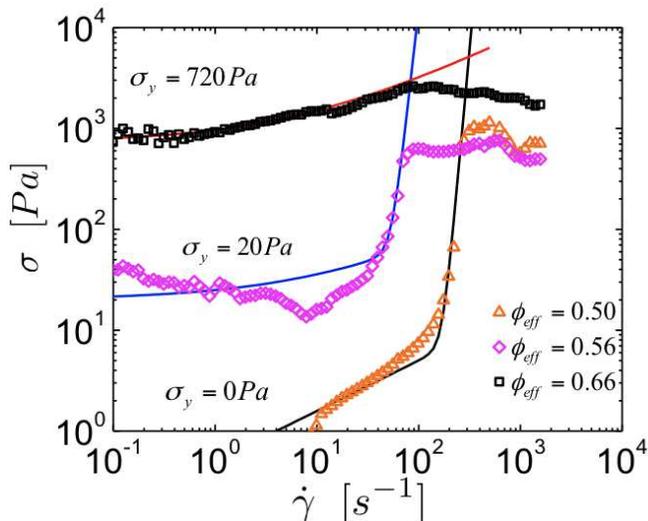}
\vspace{-0.8cm}
\caption{(color online)
The measured stress as a function of shear rate (symbols) and the extended Hershel-Bulkley curves (lines) \cite{jaegernat}.
} \label{fig3}
\end{figure}
For the shortest $\tau_{1}$, a modest shear-thinning is followed
by a Newtonian plateau. Upon increasing $\tau_{1}$, the viscosity
curves are shifted to higher values and a steep shear-thickening
appears: the rheological behavior of our system upon increasing
$\tau_{1}$ is qualitatively the same as the one of
dense non-Brownian suspensions upon increasing volume fraction
\cite{jaegerPRL, bibette}.
Note that flow inhomogeneity may occur in this type of measurements
\cite{fall, bonnoit} and we cannot exclude its presence here
(although MRI studies in a very similar system \cite{moller} detected shear
banding only for $\dot{\gamma}<60s^{-1}$). Thus
Fig.\ref{fig2} gives only macroscopic rheological information,
possibly dependent on the geometry.
A shear thinning followed by a shear thickening
regime has also been observed in some gelling colloidal suspensions
\cite{osuji}, where it has been explained in terms of an increase of
the clusters number density due to break-up. Here we have made sure
that the shear-sweep is done at $\dot{\gamma}$ lower than the value
needed for breaking up the clusters, therefore we rather
associate the shear thickening to the formation of inter-clusters
bonds, possibly also promoted by hydrodynamic interactions
\cite{wagner} and/or from a residual inter-cluster attraction:
the shear rate is too low to break the
pre-formed aggregates but can promote, instead, the formation of
further bonds among them.
At the largest $\phi_{eff}$ (i.e., $\tau_{1}$) the
shear-thickening is no longer visible.
The behavior of the stress intensity $\sigma$
as a function of $\dot{\gamma}$ plotted in Fig.\ref{fig3} sheds some light on this rheological behavior.
For large $\phi_{eff}$ the curves can be well described with an extended Herschel-Bulkley type of behavior,
which accounts for shear-thickening \cite{jaegernat},
$\sigma(\dot{\gamma})=\sigma_{y}+a_{1}\dot{\gamma}^{1/2}+a_{2}\dot{\gamma}^{1/\epsilon}$ where
$\sigma_{y}$ is the yield stress, $a_{1}$ and $a_{2}$ are parameters depending on $\phi_{eff}$ and
$\epsilon < 0.1$ is typically obtained for the steep shear
thickening of densely packed non-Brownian suspensions \cite{jaegernat}.
This indicates that the onset of the yield stress occurs in the range of
effective volume fractions $0.50< \phi_{eff}<0.56$. At $\phi_{eff}\simeq 0.66$, i.e. the
largest value obtained here, the yield stress has now a value close to the upper stress
limit of the shear thickening and this makes the shear thickening no longer
visible. The same phenomenon has been observed indeed in
densely packed non-Brownian suspensions. Hence, in a dilute colloidal
system at a solid fraction $\phi=0.21$ we get
the whole rich phenomenology observed in non-Brownian suspensions upon varying the solid fraction in a broad range.

The emergence of a yield stress at sufficiently high $\phi_{eff}$
raises the question
of how a fully solid state may form.
Upon flow cessation after each pre-shearing time $\tau_{1}$,
we have also performed
dynamic frequency sweeps. The results are reported in
Fig.\ref{fig4}.
\begin{figure}
\includegraphics[width=1.0\linewidth]{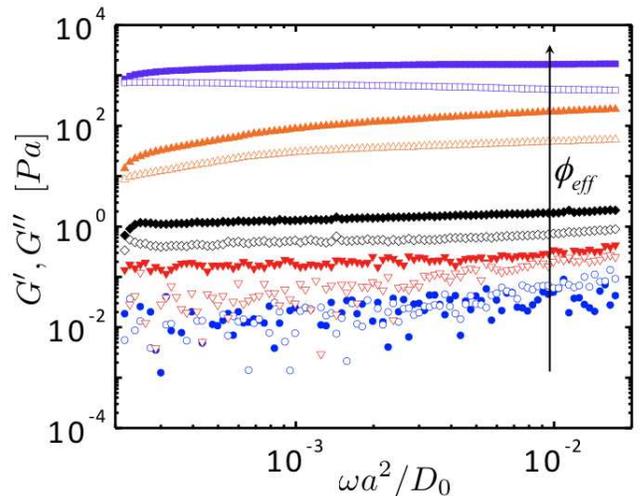}
\vspace{-0.8cm}
\caption{(color online)
$G'$ (filled symbols) and $G''$ (open symbols) measured after
$\tau_{1}= 4054 s, 6350s, 7400 s$ and $7480s$ (from bottom to top)
as a
function of the frequency $\omega$ in units of the diffusion time in
the dilute limit, $\tau_{0}=a^{2}/D_{0}$.
} \label{fig4}
\end{figure}
At the lowest $\tau_{1}$ considered (which is however close to the
point where the high-shear viscosity starts to rise dramatically)
the system still behaves liquid-like as the loss modulus $G''$ is
always slightly larger than the elastic modulus $G'$. Upon
increasing $\tau_{1}$ and hence $\phi_{eff}$, $G'$ takes over with
respect to $G''$ and the ratio $G'/G''$ becomes larger with
$\tau_{1}$, while remaining mostly constant with the frequency.
This is a behavior typically observed in colloidal gels
\cite{larson,mackley,laurati_trappe} and suggests that, for the
longest $\tau_{1}$, upon flow cessation the clusters are not only
densely packed but can also connect into a spanning, stress-bearing
structure \cite{alessio3}. During the shearing, in fact, the aggregates are
maintained in a fluid state by the imposed high shear rate, whereas upon cessation of
flow the system is subjected to a rapidly decreasing stress until a quiescent
state is reached. Hence inter-aggregates connections can gradually
form and become permanent as the zero
stress is reached: provided that $\phi_{eff}$ is high enough, a cohesive
solid random packing of aggregates will form.
To test this picture, we have calculated the elastic modulus of this disordered solid
using the approach derived in \cite{alessio3}, giving a quantitative
estimate of the contribution to elasticity of inter-aggregates connections
for an amorphous close-packed assembly of aggregates of cohesive particles:
\begin{equation}
G'=(2/5\pi)\tilde{\kappa}\phi_{eff}\tilde{z}R_{g}^{-1}
\label{ela}
\end{equation}
where $\tilde{z}$ is the aggregate average coordination number,
$R_{g}$ is their linear size and $\tilde{\kappa}$ is the stiffness of the
inter-aggregate connections formed upon flow cessation.
We use this result to estimate the elastic modulus of our aggregate gel formed
upon flow cessation at $\phi_{eff}=0.66$, corresponding to the largest effective cluster packing fraction
and the most rigid state observed,
where the affine assumption
underlying Eq.(\ref{ela}) is reasonably applicable.
The irregular aggregates
morphology, visible also by optical microscopy (inset of Fig.\ref{fig1}),
indicates that each connection
is actually composed by many colloidal bonds ({\it grainy} contacts).
In order to estimate the number of such bonds, we consider that for a
fractal aggregate of radius $R_{g}$ and fractal dimension $d_{f}$,
the total number of particles is $N_{c}= (R_{g}/a)^{d_{f}}$, $a$ being
the particle radius. Hence $a (\mathrm{d}N_{c}/\mathrm{d}R_{g})= d_{f}/a N_{c}^{(d_{f}-1)/d_{f}}$
gives the number of particles added to the outermost layer of an aggregate of radius $R_{g}$.
Using $a \simeq 60nm$, $R_{g} \simeq 35 \pm 3 \mu m$ and $d_{f} \simeq 2.7$ for our system,
we get the number of particles on the surface of the aggregates
$\simeq 136\cdot 10^3$.
With an average number of $7$ grainy contacts per aggregate, as for densely packed
spheroidal objects \cite{chaikin}, we can
estimate that each of them involves $n\simeq 19.4 \cdot 10^3$ particles,
consistent with the value $n\simeq 21 \cdot 10^3$
obtained measuring the contact area from optical micrographs.
On this basis, we estimate $\tilde{\kappa}=n \kappa$, where
$\kappa=(\partial^{2} U_{DLVO}/\partial r^{2})_{r=r_{min}}\simeq
2\cdot10^{-5}N/m$.
Hence we finally obtain for the elastic modulus of our dense {\it gel}
formed upon flow cessation, $G' \simeq 760 Pa$, consistent with the value
measured experimentally $G' \simeq 843 Pa$ at the largest $\phi_{eff}$ (see Fig.\ref{fig4}).

{\it Conclusions.} Our experiments
rationalize the shear-induced solidification
of a dilute, stable colloidal suspension for the first time in a fully
controlled way. Large aggregates of a typical size
are continuously generated under shear and behave hydrodynamically like non-Brownian hard-spheroids.
Varying the shearing time leads to the same rheological response
of dense non-Brownian suspensions upon varying the solid fraction.
Upon flow cessation, these aggregates can
eventually form \emph{cohesive} random packings where each
inter-aggregate bond involve a large number of colloidal bonds. Such
solidification mechanism is thus a hybrid between colloidal gelation
\cite{delgado} and the packing-driven jamming \cite{cates} of
non-Brownian suspensions (pastes, slurries). This scenario
gives a novel insight into the complexity of the shear-induced
solidification of colloidal dispersions of practical relevance.

{\it Aknowledgements.}
This work was supported by SNSF (Grant No. 200020-126487/1).
EDG is supported by the SNSF (Grant No. PP002\_126483/1).




\title{{Supplementary Material}\\
Shear-driven solidification of dilute colloidal suspensions}
\author{Alessio Zaccone $^{1}$, Daniele Gentili $^{1}$, Hua Wu $^{1}$, Massimo
Morbidelli$^{1}$, and Emanuela Del Gado$^{2}$}
\affiliation{${}^1$Chemistry and
Applied Biosciences, ETH Z\"urich, Switzerland}
\affiliation{${}^2$ Microstructure and Rheology, Institute for Building Materials, ETH Z\"urich, Switzerland}
\date{\today}
\maketitle

\section*{Laser Light Scattering (LLS) analysis}
\begin{figure}[h]
\includegraphics[width=1.0\linewidth]{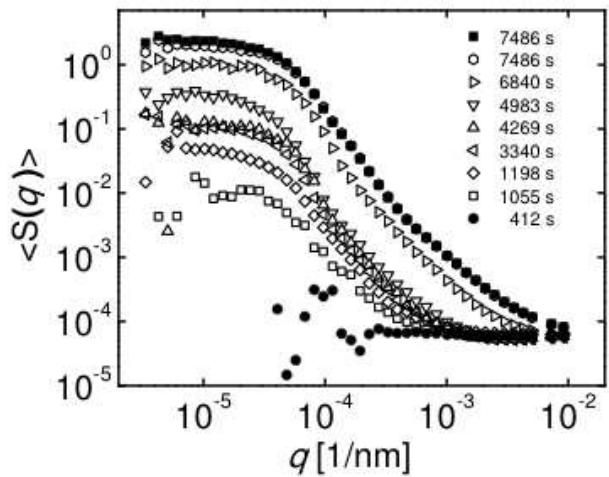}
\caption{(color online). The scattered intensity of the dilute system, $I(q)$, divided by the independently measured form factor of the latex spheres, $P(q)$, yields the average structure factor of the aggregates $<S(q)>=I(q)/P(q)$ \cite{warren}, for increasing pre-shearing time $\tau_{1}$ (from bottom to top).} \label{figA1}
\end{figure}

\begin{figure}[h]
\includegraphics[width=1.0\linewidth]{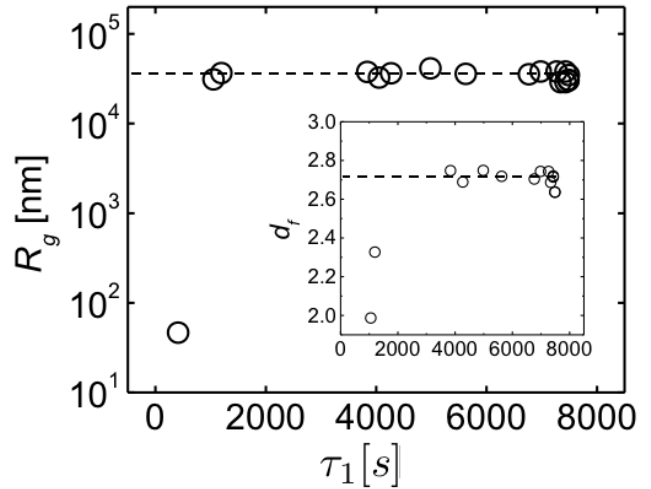}
\caption{(color online). Main frame: Average radius of gyration $R_{g}$
of the clusters (from the Guinier plot) as a function of $\tau_{1}$.
Inset (left): Average fractal dimension of the clusters $d_{f}$ as a function of $\tau_{1}$.} \label{figA2}
\end{figure}

After each pre-shearing, made at constant shear rate 
$\dot{\gamma}=1700s^{-1}$ for a duration $\tau_{1}$, we
used LLS to investigate the structure of the suspension 
in terms of the structure factor which is shown in Fig.\ref{figA1}. 
Immediately after sampling the rheometer, the samples were diluted to such an extent that multiple light scattering does not affect the measurement and such that the aggregates behave as noninteracting scatterers. At the shortest $\tau_{1}$ considered (i.e., $\tau_{1}<1000s$),
the aggregate structure factor is flat, i.e. the amount of aggregates is negligible \cite{zacJPCB}. 
Upon increasing $\tau_{1}$, the structure factor develops a plateau at low wave-vector $q$,
indicative of large clusters, followed by the power-law regime of fractal objects \cite{dimon}. Up to $\tau_{1}=4893s$, the power-law regime is followed by a flat tail which indicates that the
system is still dominated by {\it free} primary colloidal spheres coexisting with a few very large clusters and negligible amounts of clusters of intermediate size. Confirmation of this
bimodal size distribution of the aggregates comes from the analysis
of the samples after filtration with a $5\mu m$-pore size filter
which allows us to remove the large clusters from the sample. The
filtered sample is then analyzed by LLS and found to be
composed only by oligomers (monomers, dimers, and trimers)
\cite{wu2,zacthesis}.
 The gyration radius of the large clusters has been estimated from the scattering curves at different shearing times through the Guinier plot and is reported in the main frame of Fig.\ref{figA2}.
The observed plateau in the cluster linear size clearly indicates
that the aggregation under shear rapidly leads to a maximal cluster
size $R_{g}\simeq35 \pm 3\mu m$. The Guinier analysis in this regime is still applicable, as the lowest scattering angle of our instrument is $0.0145^{\texttt{o}}$, i.e. well below the limit of $0.1^{\texttt{o}}$ corresponding to the typical value of $R_{g}$. From the power-law regime of the scattering curves we also evaluate the fractal dimension $d_{f}$ of the large clusters as a
function of the shearing time (inset of Fig.\ref{figA2}). The
time evolution of $d_{f}$ also rapidly reaches a plateau at $d_{f}\simeq
2.7$. Such a fairly high fractal dimension is typical of the
shear-driven aggregation mechanism \cite{warren} and also results
from the restructuring induced by flow stresses and breakage events
\cite{becker}. 
\begin{figure}[h]
\includegraphics[width=1.0\linewidth]{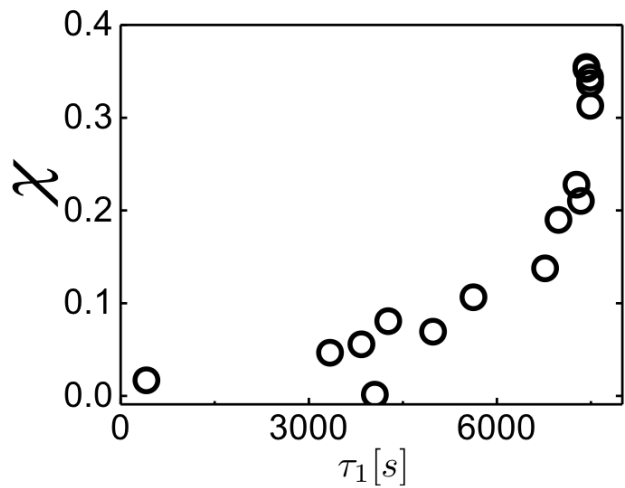}
\caption{(color online). 
Fraction $\chi$ of primary colloidal particles converted to large aggregates at constant shear rate
($\dot{\gamma}=1700s^{-1}$) as a function of the pre-shearing time $\tau_{1}$.} 
\label{figA3}
\end{figure}
Finally we have 
measured the fraction $\chi(t)$ of
primary colloidal particles converted to clusters and found that it
steeply increases with the pre-seharing time as shown in Fig.\ref{figA3}. The
fact that the size and morphology of the clusters remain constant
with time indicates that under shear the system is constantly
generating new clusters with approximately the same size ($\simeq35
\pm 5 \mu m$ of radius) and $d_{f}\simeq2.7 \pm 0.1$. The fast
increase in the number density of clusters which are fractal and
porous is associated with a rapid increase of their effective
packing fraction.

\end{document}